\documentclass[12pt]{article}

\usepackage{amssymb,amsmath,mathrsfs,epstopdf,slashed,color}

\usepackage{ifpdf}
\ifpdf
\usepackage{graphicx}
\usepackage{hyperref}
\usepackage{cite}
\else
\usepackage[dvipdfmx]{graphicx}
\usepackage[dvipdfmx]{hyperref}
\usepackage[numbers]{natbib}
\usepackage{hypernat}
\fi

\allowdisplaybreaks[1]

\makeatletter

\@addtoreset{equation}{section}
\makeatother

\DeclareMathOperator*{\re}{Re \,}
\DeclareMathOperator*{\im}{Im \,}

\title{\bf Comments on Racetrack K\"ahler Uplift}
\author{Yoske Sumitomo, S.-H. Henry Tye, and Sam S.C. Wong}

\begin{document}

\begin{titlepage}

\setcounter{page}{0}
  
\begin{flushright}
 \small
 \normalsize
\end{flushright}

\vskip 3cm
\begin{center}

{\Large \bf Statistical Distribution of the Vacuum Energy Density \\
 in Racetrack K\"ahler Uplift Models in String Theory}  

\vskip 2cm
  
{\large Yoske Sumitomo${}^1$, S.-H. Henry Tye${}^{1,2}$, and Sam S.C. Wong${}^1$}
 
 \vskip 0.6cm

 ${}^1$ Institute for Advanced Study, Hong Kong University of Science and Technology, Hong Kong\\
 ${}^2$ Laboratory for Elementary-Particle Physics, Cornell University, Ithaca, NY 14853, USA

 \vskip 0.4cm

Email: \href{mailto:yoske@ust.hk, iastye@ust.hk, scswong@ust.hk}{yoske at ust.hk, iastye at ust.hk, scswong at ust.hk}

\vskip 1.0cm
  
\abstract{\normalsize
We study a racetrack model in the presence of the leading $\alpha'$-correction in flux compactification in Type IIB string theory, for the purpose of getting conceivable de-Sitter vacua in the large compactified volume approximation.
Unlike the K\"ahler Uplift model studied previously, the $\alpha'$-correction is more controllable for the meta-stable de-Sitter vacua in the racetrack case since the constraint on the compactified volume size is very much relaxed.
 We find that the vacuum energy density $\Lambda$ for de-Sitter vacua approaches zero exponentially as the volume grows.
We also analyze properties of the probability distribution of $\Lambda$ in this class of models.
As in other cases studied earlier, the probability distribution again peaks sharply at $\Lambda=0$. We also study the Racetrack K\"ahler Uplift model in the Swiss-Cheese type model.}
  
\vspace{1cm}
\begin{flushleft}
 \today
\end{flushleft}
 
\end{center}
\end{titlepage}

\setcounter{page}{1}
\setcounter{footnote}{0}

\tableofcontents

\parskip=5pt

\section{Introduction}

Recent cosmological data strongly suggests that our universe has a vanishingly small positive cosmological constant as the dark energy,
\begin{equation}
\Lambda \sim 10^{-122}M_P^4
 \label{L1}
\end{equation}
where $M_P$ is the Planck mass \cite{Weinberg:1987dv,Bennett:2012fp,Riess:1998cb,Perlmutter:1998np} (and references therein). Arguably, its smallness is one of the biggest puzzles in modern fundamental physics. By now, it looks hopeless to search for a natural reason for its smallness within four-dimensional quantum field theory.

A typical flux compactification in string theory involves many moduli and three-form field strengths with quantized fluxes (see the review \cite{Douglas:2006es}). Together with the quantized fluxes, the moduli and their dynamics describe the string theory landscape. Since string theory has many possible vacuum solutions (i.e., leading to the so called cosmic landscape), it is argued that the spacing $\delta \Lambda$ can be exponentially small and possible values of $\Lambda$ can have large ranges. As a result, such a small $\Lambda$ (\ref{L1}) can easily be that for one of the solutions \cite{Bousso:2000xa}. However, this alone does not explain why nature picks such a very small $\Lambda$, instead of a value closer to the string or Planck scale.
 
Recently we proposed that a combination of string theory dynamics together and some basic probability theory may provide a natural order-of-magnitude explanation to this naturalness puzzle \cite{Sumitomo:2012wa,Sumitomo:2012vx,Sumitomo:2012cf}.
The basic idea is quite simple.
Suppose we can determine the four-dimensional low energy supergravity effective potential $V$ for the vacua coming from some flux compactification in string theory. To be specific, let us consider only $3$-form field strengths $F^I_3$ wrapping the three-cycles in a Calabi-Yau like manifold.
(Note that these are dual to the four-form field strengths in 4 dimensional space-time.) 
So we have $V( F^I_{3}, \phi_j) \rightarrow V(n_i, \phi_j)$, ($i=1,2,...,N, \ j=1,2,..., K$)
where the flux quantization property of the $3$-form field strengths $F_3$s allow us to rewrite $V$ as a function of  the quantized values $n_i$ of the fluxes present and $\phi_j$ are the complex moduli describing the size and shape of the compactified manifold as well as the coupling. There are barriers between different  sets of flux values. For example, there is a (finite height) barrier between $n_1$ and $n_1-1$, where tunneling between $V(n_1,n_2, ..., n_N, \phi_j)$ and $V(n_1-1, n_2,..., n_N, \phi_j)$ may be achieved by brane-flux annihilation \cite{Bousso:2000xa}.

For a given set of ${n_i}$, we can solve $V(n_i, \phi_j)$ for its meta-stable (classically stable) vacuum solutions via finding the values $\phi_{j, {\rm min}} (n_i)$ at each solution and determine its vacuum energy $\Lambda=\Lambda (n_i)$.  Collecting all such solutions, we can next find the probability distribution $P(\Lambda)$ of $\Lambda$ of these meta-stable solutions as we sweep through all the flux numbers $n_i$. Since a typical $n_i$ can take a large range of integer values, we may simply treat each $n_i$ as a random variable $b_i$ with some uniform distribution $P_i(b_i)$ and find the properties of $P(\Lambda)$.

Let us focus on Type IIB string theories with known de Sitter ($dS$) vacuum solutions. It turns out  that the string theory dynamics (i.e., the resulting functional form of $\Lambda (b_i)$) together with simple probability theory typically yields a $P(\Lambda)$ that peaks (i.e., diverges) at  $\Lambda=0$. In fact, this peaking at $\Lambda=0$ behavior is relatively insensitive to the details of the
smooth distributions $P_i(b_i)$ and  becomes more divergent as the number of moduli/fluxes increases. However, this divergence is always mild enough so $P(\Lambda)$ can be properly normalized,
i.e., $\int P(\Lambda) d \Lambda=1$, henceforth implying that the probability at exactly $\Lambda=0$ will remain exactly zero.
Since the number of moduli as well as cycles that fluxes can wrap over in a typical known flux compactification is of order ${\cal O}(100)$, a vanishingly small but non-zero $\Lambda$ appears to be statistically preferred. In fact, it is not hard to find the likely value of $\Lambda$ at comparable magnitudes as the observed value (\ref{L1}).

Although the overall emerging picture is encouraging, there are many open questions that need to be more fully addressed. In this paper, we like to study an important issue in \cite{Sumitomo:2012wa,Sumitomo:2012vx}.  In the K\"ahler Uplift model with the leading order $\alpha'$-correction \cite{Balasubramanian:2004uy,Westphal:2006tn,Rummel:2011cd,deAlwis:2011dp} in models similar to the Large Volume Scenario \cite{Balasubramanian:2005zx}, the large compactification volume approximation assumed is only moderately satisfied {\it a posteriori} for the SUSY breaking meta-stable solutions around $\Lambda=0$. Since the large volume approximation works well in the presence of this and other $\alpha'$-corrections as well as the stringy loop corrections \cite{Berg:2005ja,vonGersdorff:2005bf,Berg:2007wt,Cicoli:2007xp,Cicoli:2008va,Anguelova:2010ed}, the constraint on the volume size leads to concerns on the validity of the approximation.

The K\"ahler Uplift model studied has a single non-perturbative term in the superpotential $W$. To relax the constraint on the volume size, we generalize the model to include two non-perturbative terms in $W$, i.e., the {\it racetrack} model.
 (We like to point out that this has been briefly studied in {\cite{Westphal:2005yz,deAlwis:2011dp}}.)
Owing to this racetrack property, we find that the model admits solutions with
a large adjustable volume.

Interestingly, in this Racetrack K\"ahler Uplift model, the stability condition for both the real and imaginary sectors requires that the minima of the potential $V$ always exist for $\Lambda \ge 0$  at large volumes.
Further, the cosmological constant $\Lambda$ is naturally exponentially suppressed as a function of the volume size, and the resultant probability distribution $P(\Lambda)$ for $\Lambda$ gets a sharply peaked behavior toward $\Lambda \rightarrow  0$, which can be highly diverging.
This peaked behavior of $P(\Lambda)$ is much sharper than that of the previous K\"ahler Uplift model with a single non-perturbative term studied in \cite{Sumitomo:2012wa,Sumitomo:2012vx}.

The paper proceeds as follows : The racetrack model is introduced and reviewed in section \ref{sec:racetr-model-prop}. Among the possible solutions for $dS$ vacua, we focus on the set that allows large volumes that are not bounded from above. We contrast this with the single non-perturbative term model studied earlier. Although the range of $\Lambda$ may be unbounded from above, these $dS$ vacua are forced to have an exponentially small $\Lambda$ in the large volume limit. In section \ref{sec:prob-distr-racetr}, we present the probability distribution $P(\Lambda)$ for these $dS$ vacua, which peaks sharply at  $\Lambda=0$. In section \ref{sec:swiss-cheese-type}, we extend the racetrack potential to the Swiss-Cheese type model. Section \ref{sec:discussions-} contains the discussions and remarks.
Some details are relegated to appendix \ref{sec:case-non-zero}.

\section{A racetrack model and property \label{sec:racetr-model-prop}}

\subsection{Background}

We have focused so far on Type IIB models.
It is important to comment on the difference between type IIB and IIA models with respect to the moduli stabilization. The moduli are four-dimensional light scalar fields parametrizing the geometric size and shape (deformation) of the compact six-dimensional internal spaces (as well as the dilaton-axion mode) in string theory needed to describe our effectively four-dimensional universe.
The moduli stabilization in type IIA is typically very difficult to achieve since we have to stabilize the entire set of moduli simultaneously due to the absence of hierarchical structures.
If we have no specific structure in the potential, we may expect that the mass (squared) matrix is given rather randomly at $dS$ extremal points.
Then the probability that all eigenvalues of the random mass matrix are semi-positive (required for meta-stability)  is described by  a Gaussian suppressed function of the number of moduli \cite{Chen:2011ac,Bachlechner:2012at} (see also \cite{Marsh:2011aa}).
Since we may expect typically ${\cal O}(100)$ of moduli fields, it is clear why type IIA stabilization is so difficult to find.
On the other hand, we have the {\it no-scale} structure in type IIB; so the K\"ahler sector can be considered separately from the complex structure and the dilaton sector, which are stabilized at higher scales.
As a result, the number of moduli to be simultaneously stabilized is drastically reduced.
This hierarchical structure holds even if we introduce non-perturbative terms and $\alpha'$-corrections accordingly, which weakly break the no-scale structure so that a non-trivial potential is generated.
Some models have few K\"ahler moduli and large number of complex structure moduli.
Recently it is suggested that the large number of complex structure moduli helps to enhance the hierarchical structure \cite{Sumitomo:2012cf}.
So the IIB models are well-motivated to achieve the moduli stabilization with positive cosmological constant.

As we just pointed out, a corner of the string theory landscape where moduli stabilization can be addressed explicitly is type IIB compactified on orientifolded Calabi-Yau three-folds. The four-dimensional effective action of the geometric moduli is given by a
${\cal N}=1$ supergravity theory of a set of chiral multiplets consisting of the dilaton-axion $S$, $h^{1,1}$ number of K\"ahler moduli $T_i$, and $h^{2,1}$ number of complex structure moduli $U_i$.
The past decade saw some progress for the stabilization of $S$ and $U_i$ from the use of quantized fluxes of three form field strength of the Ramond-Ramond (RR) type $F_3=dC_2$ and Neveu-Schwarz (NS-NS) type $H_3=dB_2$, which form a $SL(2,Z)_S$ covariant three-form: $G_3 =F_3 - iS H_3$. 
They wrap three-cycles inside the manifold. Their 10-dimensional duals are $7$-form fields wrapping dual three-cycles, which result in effective four-form constant quantized field strengths in four-dimensional space-time. The flux stabilization procedure operates supersymmetrically at a high scale. The K\"ahler moduli are typically stabilized at a parametrically lower scale than the complex structure moduli via perturbative or non-perturbative interactions. We set $M_P = 1$ throughout this paper.

To be specific, we consider a racetrack model defined by the K\"ahler potential $K$ and the superpotential $W$,
\begin{equation}
 \begin{split}
  K=& - 2 \ln \left({\cal V} + {\xi \over 2}\right) - \ln \left(S + \bar{S} \right) - \ln \left(-i \int \bar{\Omega} \wedge \Omega \right),\\
  {\cal V}=& (T + \bar{T})^{3/2}, \quad
  \xi = - {\zeta(3) \over 4 \sqrt{2} (2\pi)^3} \chi (M) \left(S + \bar{S} \right)^{3/2} \sim -8.57 \times 10^{-4} {\chi} \left(S + \bar{S} \right)^{3/2}, \\
  W=& W_0 (U_i, S) + W_{\rm NP}, \quad   W_0 (U_i, S)  =  \sum_{cycles}  \int G_3 \wedge \Omega =C_1(U_i) + iS C_2(U_i), \\
   V =& e^{K} \left(K^{I \bar{J}} D_I W D_{\bar{J}} {\overline W} - 3\left|W \right|^2\right),
 \end{split}
 \label{fullpotential setup}
\end{equation}
where $\xi$ is the term coming from ${\cal O} (\alpha'^3)$ corrections to SUGRA \cite{Becker:2002nn}, and shows up as an uplifting term in the potential \cite{Balasubramanian:2004uy,Westphal:2006tn,Rummel:2011cd}. The non-perturbative term $W_{\rm NP}$ will be specified below, it is expected to be small compared to the tree-level flux contribution.
Here the dimensionless compactification volume ${\cal V} = {{\rm vol}/ \alpha'^3}$ is measured in string units.
We see that the holomorphic $3$-form $\Omega$ depends on the complex structure moduli $U_i$. 

A simplified version of $W_0(U_i, S)$ has been discussed in earlier works \cite{Sumitomo:2012vx}, where one finds the supersymmetric $W_0$ and then the ratio of the median value and the average value of the magnitude of $\Lambda$, $|\hat \Lambda| / \left< |\Lambda| \right>$, which tends to decrease exponentially as the number of $U_i$ increases. Since the stabilization of $U_i$ and $S$ are assumed to take place at a higher scale than that of the K\"ahler moduli, and this part of the analysis is very similar to the earlier work, we shall simply assume a value for $W_0$ in this paper and focus on the dynamics of the K\"ahler moduli.

\subsection{The racetrack model}

The non-perturbative terms in the superpotential $W$ as in \cite{Kachru:2003aw} (see also \cite{Escoda:2003fa,BlancoPillado:2004ns,Kallosh:2004yh}) are crucial in the K\"ahler moduli stabilization. 
Compared to our earlier work \cite{Sumitomo:2012wa,Sumitomo:2012vx}, the main new feature here is the presence of the new term $B e^{-b T}$ in $W$. Together with the other terms in $W$, this forms the so called ``racetrack''.
 To focus on this feature, let us assume that dilaton and complex structure moduli are stabilized supersymmetrically $D_S W = \partial_SW +(\partial_SK)W=0, D_{U_i} W = 0$ at some higher energy scale, so $W$ is reduced to
 \begin{equation}
W= W_0 + W_{\rm NP} = W_0 + A e^{- a T} + B e^{-b T} 
\end{equation}
where the coefficients $a$ and $b$ are taken to be positive real.
The potential $V$ for the K\"ahler modulus $T$ becomes
\begin{equation}
 \begin{split}
  V =& e^{K} \left[ K^{T \bar{T}} \left( \partial_T W \partial_{\bar{T}} \bar{W} + \partial_T W (\bar{W} \partial_{\bar{T}} K) + \partial_{\bar{T}} \bar{W} (W \partial_T K)  \right)  + 3 \xi {\xi^2 + 7 \xi {\cal V} + {\cal V}^2 \over ({\cal V}- \xi) (\xi + 2 {\cal V})^2} \left|W\right|^2 \right].
 \end{split}
 \label{fullpotential}
\end{equation}
In this paper, we are mainly interested in meta-stable $dS$ vacua achieved with the uplifting $\alpha'$-correction term $\xi$ in the K\"ahler potential.

Note that a solution of this type was found and used for an inflationary universe analysis in \cite{Westphal:2005yz}.
Here we are interested in more detail properties and systematic understandings of this class of solutions, including allowed region of parameters for $dS$ uplift, such that we can analyze also the property of distribution of vacua.
We also like to point out that the model has been briefly studied in {\cite{deAlwis:2011dp}}, but in a different parameter region.

Before proceeding, it would be interesting to see what happens for supersymmetric vacua obtained before turning on $\alpha'$-correction.
If the supersymmetric condition $D_T W |_{\xi=0} = 0$ holds before turning on $\alpha'$-correction, the potential becomes
\begin{equation}
 \begin{split}
  V = e^{K} \left( |D_T W|^2 - 3 |W|^2 \right) \sim 3 e^{K_{\xi=0}} |W|^2 \left( -1 + {\xi \over {\cal V}}\right) 
 \end{split}
\end{equation}
where in the last equation, we kept up to the leading order of $\xi/{\cal V}$.
We see immediately that $dS$ vacua $V|_{\rm min} > 0$ by the uplifting from SUSY $AdS$ requires ${\xi/{\cal V}} \gtrsim {\cal O} (1)$.
This clearly violates our assumption that $\alpha'$-correction is under control in type IIB SUGRA approximation.
Thus we need SUSY breaking vacua before the uplift by the leading order $\alpha'$-correction.
Note that the leading term of $|D_T W|^2$ starts with ${\cal O} (\xi^2)$ due to the supersymmetric condition.\footnote {Although our interest throughout this paper is for the uplift by the leading order $\alpha'$-correction, it is worth commenting that the SUSY vacua can be uplifted to $dS$ by introducing an explicit SUSY breaking term like D3-$\overline{\rm D3}$ pairs contribution considered in \cite{Kachru:2003aw, Kachru:2002gs}.}

In the large volume situation where $\xi/{\cal V}  \ll 1$, which is of interest here, the potential may be approximated to
\begin{equation}
 \begin{split}
  &V \simeq \left(-{a^3 A \, W_0\,  \over 2}\right) \lambda (x,y), \\
  & \lambda (x,y) =  - {e^{-x} \over x^2} \cos y - {\beta \over z} {e^{-\beta x} \over x^2} \cos (\beta y) + {\hat{C} \over x^{9/2}},\\
  &x = a t, \quad y=a \tau, \quad   z = A/B, \quad \beta = b/a, \quad {\hat{C}} = -{3 a^{3/2} W_0 \, \xi \over 32 \sqrt{2} A}, 
 \end{split}
 \label{approximated potential}
\end{equation}
where we define $t = \re T,\  \tau=\im T$, and treat $W_0, A, B$ as real parameters for simplicity. Note that we have dropped
the $e^{-2x}$, $e^{-2 \beta x}$ and $e^{-x-\beta x}$ terms as well as ${\cal O}\left(({\xi \over {\cal V}})^2\right)$ terms  in anticipation of a large $x$ or, equivalently, a large $\cal V$.
This approximation is essentially the same as that used for the case with a single non-perturbative term \cite{Rummel:2011cd,Sumitomo:2012wa,Louis:2012nb,Sumitomo:2012vx,deAlwis:2011dp}.
Here, we are interested in the parameter region where we can have a large dimensionless volume ${\cal V} \gg 1$. Once we have the desired solution, we can check {\it a posteriori} the validity of this approximation.

Note that this model (\ref{approximated potential}) reduces to the single non-perturbative term model if either (1) $\beta=1$, where $A \rightarrow A+B$, or (2) $B=0$ ($z^{-1}=0$). 
This single non-perturbative term case has been studied carefully  \cite{Rummel:2011cd} (where $C=9/2\, \hat{C}$ is used) and a meta-stable $dS$ vacuum is obtained (with $\tau=0$) only if 
\begin{equation}
\label{single term}
  0.811 \le {\hat{C}} \le 0.864,    \quad  5/2 \le x \le 3.11
\end{equation}
where the lower $x$ value corresponds to a Minkowski vacuum and the upper bound indicates the vanishing of the modulus mass. This also requires $W_0/A$ (and so $W_0A$) to be negative.
The validity of the large volume approximation requires
$${\xi/ 2 {\cal V}} \sim a^{3/2} \xi/12\sqrt{6} \ll 1.$$
The $\alpha'$-correction term $\xi$ is given as in (\ref{fullpotential setup}), and we do not expect this term to be tiny since we like to stay in weak coupling regime $\re S > 1$ and $-\chi \sim {\cal O} (100)$.
Although we may take the volume large by taking a smaller $a$, we do not expect this to happen naturally.
The non-perturbative term is obtained by Euclidian D3-brane or gaugino condensation on D7-branes.
For instance, $a=2\pi/N$ for $SU(N)$ gaugino condensation on D7-branes.
Recently it is analyzed that the D7-brane tadpole cancellation \cite{Collinucci:2008pf} as well as the holomorphicity of D7-branes suggests an upper bound for the maximal rank of gauge group on D7-branes \cite{Cicoli:2011qg,Louis:2012nb} (where $N \le 24$ or $a \ge 0.26$ in a specific model) .
So there exists an upper bound on $\cal V$ itself for a non-racetrack type of the K\"ahler Uplift scenario.

There are a number of meta-stable solutions to $V$ (\ref{approximated potential}):
\begin{itemize}
\item For very small $B$, the $B$ term may be treated perturbatively. Then the single term constraints (\ref{single term}) will only be slightly modified. That is, the volume $\cal V$ is still strongly constrained from above.
\item For $\beta \gg 1$, the dropped $e^{-2x}$ term will become more important than the kept $e^{-\beta x}$ term in the approximate $V$ (\ref{approximated potential}). In this case, we should restore the $e^{-2x}$ term and may treat the $B$ term perturbatively. Again, the volume $\cal V$ will be strongly constrained as before.
\item For small $\beta$,  it turns out that $\beta \ll 1$; in fact, $\beta \sim 1/x$, and so the dropped $e^{-2\beta x}$ term will be more important than the kept $e^{-x}$ term.
      Assuming we have a solution when $\beta \sim 1/x$, then the volume is constrained above by ${\cal V} \sim (N_b/2\pi)^{3/2} \lesssim 21$ where we used the bound $N_b \leq 24$ for ${\mathbb P}^4_{11169}$ \cite{Louis:2012nb} as an illustration.
\item Since we like to find solutions where $x$ may be taken large without much constraint, we shall focus on the solutions where $\beta \sim {\cal O} (1)$ and $B$ needs not be small.
\end{itemize}

\subsection{Large volume solution}

Since the $\tau \neq 0$ case at large volume has no stabilized solution, as explained in appendix \ref{sec:case-non-zero},
let us focus on the $y=a \tau=0$ case.
Here, we like to show that the stability conditions can be met at the large volume approximation for $y=0$ and $\beta \sim {\cal O}(1)$.
Interestingly, the solution with $\beta \gtrsim 1$ and large volume imply a classically stable $dS$ vacuum with an exponentially small cosmological constant $\Lambda$. 
Also we shall check that the solution of the approximated potential (\ref{approximated potential}) is well justified even in the full potential (\ref{fullpotential}).

Now we solve for the classically stable minima in the $\beta \sim {\cal O}(1)$ region.
The extremal conditions $\partial_t V = \partial_{\tau} V = 0$ may be expressed as the relations:
\begin{equation}
 \begin{split}
  {1\over z} =& e^{(\beta-1) x} {- 2 x + 5 + 9 e^{x} x^2 \lambda \over \beta (2 \beta x - 5)},\quad
  {\hat{C}} =  2 e^{-x} x^{7/2} { (\beta-1) + e^{x} (x^2 + 2 x) \lambda \over 2 \beta x -5},
 \end{split}
 \label{relations for extrema}
\end{equation}
where the rescaled potential $\lambda$ will be constrained by the stability condition.
Plugging $z, C$ in the Hessian $\partial_i \partial_j V \propto  \partial_i \partial_j \lambda$ and taking the large volume expansion $x\gg 1$, we get
\begin{equation}
 \begin{split}
\partial_x^2 \lambda |_{\rm ext} \simeq&  e^{-x} \left({\beta-1 \over x^2} - {5(\beta-1) \over 2 x^3}
    \right)
  + \lambda \left(-{9 \beta \over 2 x} - {9 \over 2 x^2}
    \right) + \cdots \\
  &= m^2 \ge 0 , \\
 \partial_y^2 \lambda|_{\rm ext} \simeq & e^{-x} \left(-{\beta-1 \over x^2} + {5 (\beta-1) \over 2 x^3}
   \right)
  + \lambda \left({9 \beta \over 2 x} + {45 \over 4 x^2}
   \right) + \cdots \\
  & = - m^2 +  \lambda \left({63 \over 4 x^2} \right) + \cdots  \ge 0,
 \end{split}
 \label{stabilityc}
\end{equation}
while the off-diagonal component  $\partial_x \partial_y V|_{\rm ext}=0$ at the $y=0$ value.
So the stability condition (positive mass squared for both $x=at$ and $y=a\tau$ at the extremum) puts a strong constraint on the value of $\lambda=- 2 V|_{\rm ext}/ a^3 A W_0$.

The stability condition (\ref{stabilityc}) in the large volume approximation yields $\beta > 1$ and
\begin{equation}
 \begin{split}
e^{-x}{2 (\beta-1) \over 9 \beta x} \left(1 - {5 (\beta +1) \over 2 \beta x} \right) \leq \lambda \lesssim e^{-x}{2 (\beta-1) \over 9 \beta x}  \left(1 - {5(\beta + 1) \over 2 \beta x } + {3 \over 2 \beta x } \cdots \right). 
 \end{split}
 \label{leading lambda}
\end{equation}
which takes the form $0 \le \lambda_{\rm min} \le \lambda \le \lambda_{\rm max}$. So we see that a positive but small $\lambda$ is guaranteed together with the large volume ${\cal V}$ and $\beta > 1$, implying that a positive cosmological constant $\Lambda =  V|_{\rm min} = \left(- a^3 A \, W_0 / 2 \right) \lambda = a^3 |A \, W_0 | \lambda/ 2$ emerges.
It is interesting that the stability condition imposes no upper bound on $x$, in contrast to the single term case (\ref{single term}). In fact, $\lambda_{\rm min} \rightarrow \lambda_{\rm max}$ as $x \rightarrow \infty$, so $\Lambda$ approaches an exponentially small positive value at the large volume ($x \rightarrow \infty$) limit,
\begin{equation}
 \Lambda \sim  \left({a^3 |A \, W_0 |} \over 2 \right) e^{-x}{2 (\beta-1) \over 9 \beta x} \left(1 - {5 (\beta +1) \over 2 \beta x} \right).
\end{equation}
As we will analyze the probability distribution $P(\Lambda)$ in the next section, a small $\Lambda$ is quite generic in the Racetrack K\"ahler Uplift model.

It is also interesting to find the lower bound on $x$. To do so, we have to start with the exact formulae for $\partial_i \partial_j \lambda|_{\rm ext}$. With that, we can easily write down the large volume limit of the constraints on $\lambda$,
\begin{equation}
 \begin{split}
 \label{exactcond}
  e^{-x} {(\beta-1) (2 \beta x -5 (\beta+1)) \over 9 \beta^2 x^2} \le \lambda \le e^{-x} {(\beta-1) (4 \beta x^2 - 10 (\beta+1) x + 35) \over 9 x (2 \beta^2 x^2 - 3 \beta x-10)}.
 \end{split}
\end{equation}
We see that $\lambda_{\rm max} \rightarrow \infty$ as its denominator factor $(2 \beta^2 x^2 - 3 \beta x-10) \rightarrow 0$. This yields the lower bound for $x$
\begin{equation}
x \ge \frac{3 + \sqrt{89}}{4 \beta}  \sim  \frac{3.11}{\beta}.
\end{equation}
As $x$ approaches this bound, the range of allowed $\lambda$ becomes infinite.
For $x$ smaller than this value, $\lambda_{\rm max} <0$. Note that this lower bound matches the upper bound (\ref{single term}) as $\beta \rightarrow 1$.

So far we have restricted the parameters in terms of $x$.
Now we estimate the solution for $x$ in terms of the parameters in the large volume limit satisfying all the conditions.
Inserting the leading order of $\lambda$ (\ref{leading lambda}) back into the extremal conditions (\ref{relations for extrema}), we have at large $x$,
\begin{equation}
 {1 \over z} \sim - {1\over \beta^3} e^{(\beta-1)x}, \quad \hat{C} \sim { 2(\beta -1) \over 9\beta} e^{-x} x^{7/2}.
  \label{extremal relation for x}
\end{equation}
If we solve the first equation for $x$, we get
\begin{equation}
 x \sim {1\over \beta-1} \ln \left({\beta^3 \over - z}\right), \quad
  \hat{C} \sim {2\over  9\beta (\beta-1)^{5/2}} \left({\beta^3 \over -z}\right)^{-1\over \beta-1} \left(\ln {\beta^3\over  -z }\right)^{7/2}.
  \label{large-volume-sol}
\end{equation}
Therefore $\beta$ close to one as well as small $|z|$ contribute to the large volume.
To obtain the large volume solution, we consider the parameter region so that $0< - \beta^{-3} z < 1$.
Plugging the large volume solution (\ref{large-volume-sol}) to the leading order of the cosmological constant $\Lambda$ obtained in (\ref{leading lambda}), the absolute values of the potential minimum is given by
\begin{equation}
\Lambda =  \left. V\right|_{\rm min} \sim \left({a^3 B \, W_0  \over 2}\right) {2 (\beta-1)^{2} \beta^{2} \over 9} {(- \beta^{-3} z)^{\beta \over \beta-1} \over -\ln (-\beta^{-3} z )}.
  \label{potmin at large volume 0}
\end{equation}
Note that $A$ in the coefficient is rewritten in terms of $z$.
Since the range of $\Lambda$ restricted by the stability analysis is quite narrow for a given value of $z$, the formula here gives the right magnitude for $\Lambda$ at the meta-stable vacua.
Next, let us see how the condition for $\hat{C}$ (\ref{large-volume-sol}) can be met.
If we have a non-trivial Euler number $\chi$, as is apparent from the formula in (\ref{fullpotential}), the smallest possible $\xi$ may be of order ${\cal O} (10^{-3})$ assuming weakly string coupling $\re S = 1/g_s >1$.
Together with a lower bound for $a$, a natural requirement to satisfy $\hat{C}$ in (\ref{extremal relation for x}) is having a small $W_0$.
So $\Lambda$ is further suppressed due to the small $W_0$ in (\ref{potmin at large volume 0}).
To make this point clear, we substitute $W_0$ using equation (\ref{extremal relation for x}) and get
\begin{equation}
 \Lambda \sim {62\sqrt{2} a^{3/2} B^2 \beta^4 \over 243 \xi \sqrt{\beta-1}} \left(-\beta^{-3} z\right)^{2\beta \over \beta-1} \left(-\ln (-\beta^{-3} z)\right)^{5/2}.
 \label{potmin at large volume}
\end{equation}
When $\beta \gtrsim 1$ and simultaneously $|z|$ is small, we have an exponential suppression in the cosmological constant.

Finally let us estimate how the $\alpha'$-correction is controllable.
Using the large volume solution (\ref{relations for extrema}) and (\ref{large-volume-sol}), we get
\begin{equation}
 {\hat{C} \over x^{3/2}} \sim  {2\over  9\beta (\beta-1)} \left({- \beta^{-3} z }\right)^{1 \over \beta-1} \left|\ln \left( -{\beta^{-3} z} \right)\right|^2.
\end{equation}
Since this uplifting term is highly suppressed as a function of $z$ and $\beta$, our approximation keeping up to the linear term for $\alpha'$-correction in (\ref{approximated potential}) works quite well.
On the other hand, the suppression ratio for the $\alpha'$-correction becomes
\begin{equation}
 {\xi \over 2 {\cal V}} \sim  {a^{3/2} \xi \over 4\sqrt{2}} {(\beta-1)^{3/2} \over |\ln (-\beta^{-3} z)|^{3/2}}.
\end{equation}
It is clear that if we have only single non-perturbative term, there is no suppression depending on $z, \beta$.
So this large volume approximation also works to make the required $\alpha'$-correction smaller such that type IIB SUGRA approximation stays valid.
The construction with the large volume makes several corrections under control, including higher $\alpha'$ or stringy loop corrections, which may scale as ${\cal O} ({\cal V}^{-10/3})$ in the potential \cite{Berg:2005ja,vonGersdorff:2005bf,Berg:2007wt,Cicoli:2007xp,Cicoli:2008va,Anguelova:2010ed} in light of the {\it extended no-scale structure} \cite{Berg:2007wt,Cicoli:2007xp}.

\begin{figure}
 \begin{center}
  \includegraphics[width=25em]{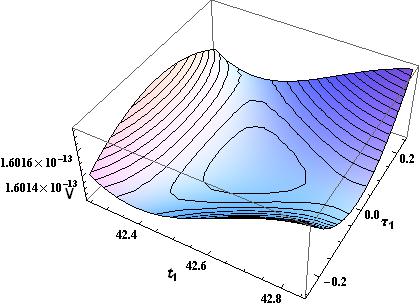}
 \end{center}
 \caption{\footnotesize The minimum in the full potential (\ref{fullpotential}) with the parameters given in (\ref{parameter set}). The minimum sits at $t=42.5, \ \tau = 0$.}
 \label{fig:fullpot-single}
\end{figure}

Here we present an example for illustration. Let us start with the following input parameters:
\begin{equation}
 \begin{split}
  W_0 = -0.223, \quad A = 1.65, \quad B = - 4.77, \quad a = {2\pi \over 15}, \quad b = {2 \pi \over 14}, \quad \xi = 3.41 \times 10^{-3}.
 \end{split}
 \label{parameter set}
\end{equation}
So the combined parameters are given by (to 3 digits) 
\begin{equation}
 \begin{split}
  z = -0.346, \quad \hat{C} = 8.28 \times 10^{-6}, \quad  \beta = {15\over 14} \simeq 1.07, \quad {-z \over \beta^3} \simeq  0.281,
 \end{split}
\end{equation}
We find the minimum at $t = 42.5, \ \tau = 0$ in the full potential (\ref{fullpotential}) as shown in figure \ref{fig:fullpot-single}, where the volume is quite large, ${\cal V} = 785$.

The solution in the full-potential (\ref{fullpotential}) with input (\ref{parameter set}) is given by 
\begin{equation}
 \begin{split}
  x^{(\rm full)} = 17.8228, \quad \Lambda= V^{(\rm full)}|_{\rm min} = 1.60122 \times 10^{-13}.
 \end{split}
\end{equation}
while the approximate potential (\ref{approximated potential}) yields
\begin{equation}
 \begin{split}
  x^{(\rm approx)} = 17.8219, \quad V^{(\rm approx)}|_{\rm min} = 1.60115 \times 10^{-13}.
 \end{split}
\end{equation}
On the other hand, the approximated analytical formulae (\ref{large-volume-sol}) and (\ref{potmin at large volume}) yield
\begin{equation}
 \begin{split}
  x^{(\rm anal)} = 17.7597, \quad V^{(\rm anal)}|_{\rm min} = 1.78635 \times 10^{-13}.
 \end{split}
\end{equation}
So we see that the approximations work quite nicely not only for the approximate potential (\ref{approximated potential}), but also for the analytic expressions.

\section{Probability distribution of Racetrack K\"ahler Uplift \label{sec:prob-distr-racetr}}

In previous section, we see that the racetrack K\"ahler Uplift model have no upper bound for the volume moduli, unlike the K\"ahler Uplift model with a single non-perturbative term.
To understand how likely a tiny cosmological constant will appear in this racetrack model, we analyze the probability distribution $P(\Lambda)$ in this section. 
The $\Lambda$ at the classically stable minimum of the potential (\ref{potmin at large volume}) is given by
\begin{equation}
 \Lambda \equiv {62\sqrt{2} a^{3/2} B^2 \beta^4 \over 243 \xi \sqrt{\beta-1}} \kappa^{2\beta \over \beta-1} \left(-\ln \kappa\right)^{5/2},  \quad
  \kappa \equiv {-z \over \beta^3}.
  \label{lambda}
\end{equation}

Now we introduce a randomness to the system.
As discussed in \cite{Sumitomo:2012wa,Sumitomo:2012vx,Sumitomo:2012cf,MartinezPedrera:2012rs,Danielsson:2012by}, when we take into account the moduli stabilization of complex structure moduli and dilaton with different values of fluxes, we expect that the different values of $W_0, A, B$ are given.
Together with the fact that there are also many types of models for complex structure moduli, corresponding to many varieties of Calabi-Yau compactifications, we have a rich enough structure of vacua in the string landscape.
To deal with all of these models is rather complicated, so we mimic this variety by just simply randomizing some parameters.
Let $W_0$ be a random parameter obeying a uniform distribution with a range that satisfies the condition for $\hat{C}$ in (\ref{leading lambda}), so the expression (\ref{lambda}) follows. To simplify the analysis, we set $B =1$ and  $\xi =1$.
Since $B$ contributes just in the coefficient and does not really touch the details of the dynamics, we set simply $B =1$ for simplicity of the arguments.
Learning from the analysis in \cite{Sumitomo:2012wa,Sumitomo:2012vx}, it is clear that randomizing $B$ will probably not diminish the divergent peak in the distribution $P(\Lambda)$ towards $\Lambda = 0$.
We also do not randomize the parameters $a, \beta$. The Euclidian D3-brane gives us the non-perturbative term with $a = 2\pi$, while the gaugino condensation on D7-branes produce the term, e.g. with  $a = 2 \pi/N$ for $SU(N)$ group.
But since we also have to satisfy the tadpole cancellations in Calabi-Yau compactification, which affects the number of the D7-branes to keep its holomorphicity \cite{Cicoli:2011qg,Louis:2012nb}.
So since there may not remain large choices for these $a, b$, we rather pick up a value for $\beta$ and set $a=1$ for simplicity.

Note that the statistical distribution of the flux vacua was considered by giving a distributed randomness to the flux quantities in \cite{Ashok:2003gk,Denef:2004ze,Denef:2004cf}.
In contrast, our interest here is to estimate the probability distribution $P(\Lambda)$ of $\Lambda$ of the flux vacua, especially in the concrete $dS$ cases so that the stabilization dynamics crucially affects the distribution.
The shape of the distribution would imply how likely we can achieve small values for $\Lambda$.

By setting $a = B = \xi = 1$ for simplicity, the probability distribution is estimated by the formula:
\begin{equation}
 \begin{split}
  P(\Lambda) = \int dz \,  P(z) \, \delta \left( \Lambda - {62 \sqrt{2} \beta^4 \over 243 \sqrt{\beta-1}} \kappa^{2\beta \over \beta-1} (-\ln \kappa)^{5/2} \right).
 \end{split}
 \label{PDF-lambda}
\end{equation}
For a fair discussion, we consider the uniformly distributed $-1 \leq z \leq 0$ with $P(z) = 1$ as a conservative choice.
Then the integration can be performed quite easily by
\begin{equation}
 P(\Lambda) = {243 (\beta-1)^{3/2} \over 32 \sqrt{2} \beta} {\kappa^{1+\beta \over 1-\beta} \over (- \ln \kappa)^{3/2} (5-5\beta - 4 \beta \ln \kappa)}.
  \label{PDF2}
\end{equation}
The range $-1 \leq z \leq 0$ is good for the large volume approximation. For instance at $\beta=1.1$, we have $x \sim -\ln \kappa/(\beta-1) \geq 2.86$, and the distribution (\ref{PDF2}) is a well-defined monotonically decreasing function of $\kappa$.

Since we would like to rewrite this as a function of $\Lambda$, we solve (\ref{lambda}) for $\kappa$ by
\begin{equation}
 \ln \kappa = {5 (\beta-1) \over 4 \beta} {\cal W}_{-1} \left(-{9 \Lambda^{2/5} \over 5 \times 2^{3/5} \beta^{3/5}  (\beta-1)^{4/5} }  \right).
\end{equation}
Here we introduce the {\it Lambert W}-function which is the solution of ${\cal W} e^{\cal W} = X$, and ${\cal W} = {\cal W}_{-1} (X)$ when ${\cal W} \leq -1$,
since our large volume approximation works if $4 \beta \ln \kappa/5(\beta-1) \sim -{4\beta x/5} \ll -1$.
Inserting this solution into (\ref{PDF2}), we get
\begin{equation}
 \begin{split}
 P(\Lambda) =& - {243 \beta^{1/2} \over 100\sqrt{10} (\beta-1)} {e^{-{5(\beta+1) \over 4 \beta} {\cal W}_{-1}} \over (-{\cal W}_{-1})^{3/2} (1+{\cal W}_{-1})}, \\
  {\cal W}_{-1} =& {\cal W}_{-1} \left(-{9 \Lambda^{2/5} \over 5 \times 2^{3/5} \beta^{3/5}  (\beta-1)^{4/5} }  \right).
  \label{PDF-final}
 \end{split}
\end{equation}

Let us expand (\ref{PDF-final}) around $\Lambda \sim 0$ to get the asymptotic behavior.
Using the expansion of ${\cal W}_{-1}$ for small $X$:
\begin{equation}
 {\cal W}_{-1} \left({X}\right) \sim \ln X - \ln (-\ln X) + \cdots,
\end{equation}
the probability distribution becomes
\begin{equation}
 P(\Lambda) \stackrel{\Lambda \rightarrow 0}{\sim} {243 \beta^{1/2} \over 16 (\beta-1)} {1 \over \Lambda^{\beta+1 \over 2 \beta}  (-\ln \Lambda)^{5/2}}.
  \label{asymptotics of PDF}
\end{equation}
So for $\beta \gtrsim 1$, we see that the diverging behavior is very peaked as  $\Lambda \rightarrow 0$.
Since $(\beta +1)/2 \beta < 1$, $P(\Lambda)$ is normalizable, i.e.,$\int P(\Lambda) d \Lambda =1$.

\begin{figure}
 \begin{center}
  \includegraphics[width=20em]{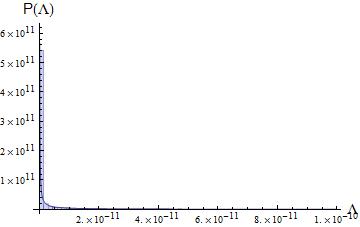}
  \includegraphics[width=20em]{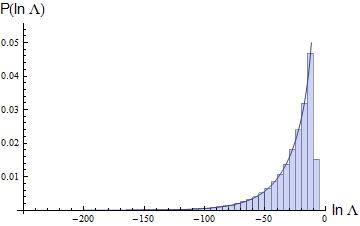}
 \end{center}
 \caption{\footnotesize The asymptotics at $\beta = 1.1$ of the analytical expression (\ref{PDF-final}) and the numerical data.}
 \label{fig:PDF-numerical}
\end{figure}

We illustrate the result in the figures of the probability distribution function of $\Lambda$ in (\ref{PDF-lambda}).
Here we again set $a = B = \xi = 1$ and uniformly distributed $-1 \leq z \leq 0$.
The analytical expression (\ref{PDF-final}) as well as the numerical histogram of (\ref{lambda}) are illustrated in figure \ref{fig:PDF-numerical} at $\beta=1.1$.
The distribution is quite sharply peaked toward $\Lambda = 0$, as estimated in (\ref{asymptotics of PDF}).

To get a better feeling of quantification of the peaking behavior, it would be better to introduce the likely value that $Y\%$ of the data fall in the value $\Lambda_Y$: $\int_0^{\Lambda_Y} d\Lambda \, P(\Lambda) = Y\%$.
Using the data obtained above at $\beta=1.1$, the likely values become
\begin{equation}
 \begin{split}
  \Lambda_{10} = 3.61 \times 10^{-24},\quad \Lambda_{50} = 7.08 \times 10^{-10}, \quad \Lambda_{80} = 4.25 \times 10^{-6}, \quad \left< \Lambda \right> = 8.90 \times 10^{-6},
 \end{split}
\end{equation}
where we also present the average value for comparison.
We see that just $10\%$ fine-tuning suggests the substantial suppression of cosmological constant.
This is nothing but because the highly sharply peaked behavior as in (\ref{asymptotics of PDF}).
Note that the $\Lambda_{50}$ is simply the median.
For comparison, we have, at $\beta=1.04$, a much sharper peaking behavior as is clear from (\ref{asymptotics of PDF}):
\begin{equation}
 \Lambda^{\beta=1.04}_{10} = 2.83 \times 10^{-54}, \ \Lambda^{\beta=1.04}_{50} = 5.47 \times 10^{-19}, \ \Lambda^{\beta=1.04}_{80} = 2.81 \times 10^{-9}, \ \left< \Lambda \right>^{\beta=1.04} = 6.36 \times 10^{-7}. 
\end{equation}

\section{Swiss-Cheese type model \label{sec:swiss-cheese-type}}

So far we have focused on a single K\"ahler modulus case.
Here we introduces multi-K\"ahler moduli and check whether the multi-moduli case is compatible with the large volume approximation in the Racetrack K\"ahler Uplift, especially with the {\it Swiss-Cheese} type of compactification.

The Swiss-Cheese type model is a class of Calabi-Yau compactification, and is used to realize the Large Volume Scenario (LVS) \cite{Balasubramanian:2005zx}.
It is clarified that there is a large variety of Swiss-Cheese type of compactification \cite{Denef:2004dm,Cicoli:2008va,Gray:2012jy}. 
In LVS, the volume is made actually quite huge.
The large volume is good to have a control of several corrections including higher $\alpha'$ or stringy loop corrections, scaling like ${\cal O} ({\cal V}^{-10/3})$ in the potential \cite{Berg:2005ja,vonGersdorff:2005bf,Berg:2007wt,Cicoli:2007xp,Cicoli:2008va,Anguelova:2010ed}.
In our analysis with the Racetrack K\"ahler Uplift, although we do not expect that the volume can be huge naturally as well as LVS, we consider that the volume may be large enough so the corrections are under control.

We focus on a two K\"ahler moduli case as a test example to investigate the multi-K\"ahler scenario.
The model is given by
\begin{equation}
 \begin{split}
  K =& - 2 \ln \left({\cal V} + {\xi \over 2} \right)+ \cdots, \\
  {\cal V} =& \left( T_1 + \bar{T}_1 \right)^{3/2} - \left( T_2 + \bar{T}_2 \right)^{3/2},\\
  W =& W_0 + A_1 e^{-a_1 T_1} + B_1 e^{-b_1 T_1} + A_2 e^{-a_2 T_2}.
 \end{split}
 \label{full-pot for two}
\end{equation}
Here we just introduce single non-perturbative term for the second modulus.
Again we assume that the complex structure moduli and dilaton are stabilized supersymmetrically, and choose the solution for imaginary modes to be $\im T_1 = \im T_2 = 0$ for simplicity.
We are interested in the parameter region which include $dS$ vacua as a result of the precise K\"ahler uplift, the potential may be approximated up to leading order of the non-perturbative term as well as the $\alpha'$-correction, by
\begin{equation}
 \begin{split}
  V =& \, e^{K} \left(|DW|^2 - 3 |W|^2 \right)\\
  \sim&\, \left(- {a_1^3 B_1 W_0 z \over 2} \right) \left(- {x_1 e^{-x_1} \over {v}^2} \cos y_1 - {\beta \over z} {x_1 e^{-\beta x_1}  \over v^2} \cos (\beta y_1) - d_2 {x_2 e^{-x_2} \over v^2} \cos y_2 + {\hat{C} \over v^3} \right),\\
  v =&\, x_1^{3/2} - \left({x_2 \over \delta_2}\right)^{3/2}, \quad
  x_2 = a_2 t_2, \quad y_2 = a_2 \tau_2, \quad d_2 = {A_2 \over A_1}, \quad \delta_2 = {a_2 \over a_1}.
 \end{split}
 \label{eff-pot for two}
\end{equation}
We analyze the dynamics of this effective potential.

First, we consider the dynamics on $y_2$ direction, where the second derivative is given by
\begin{equation}
 \partial_{y_2}^2 V = {d_2 x_2 e^{-x_2} \over v^2} \cos y_2.
\end{equation}
Together with the fact that the first derivative is proportional to $\sin y_2$, we can have a stable solution at $y_2=0$ when $d_2 >0$.
The off-diagonal component with respect to $y_2$ at the extrema $\partial_i \partial{y_2} V |_{\rm ext}$ is now trivial due to the solution $y_2=0$.
Since $y_1=0$ solution is also motivated as argued in the previous section and $x_2$ direction can not change the dynamics for $x_1, y_1$ at the large volume, so we take the solution with $y_1 = y_2 =0$.

Similarly to the previous analysis, the extremal condition $\partial_i V =0$ with $\lambda = - 2V|_{\rm ext} /a^3 B W_0 z$ can be rewritten by
\begin{equation}
 \begin{split}
  {1\over z} \sim& e^{(\beta-1) x_1} \left( - {1\over \beta^2} + {5 (\beta-1) \over 2 \beta^3 x_1} \right) + e^{\beta x} \lambda \left( {9 x_1 \over 2 \beta^2} + {45 \over 4 \beta^3} \right) + \cdots,\\
  {d_2} \sim& {e^{-x_1+x_2} \over 1-x_2} \left({3 (\beta-1) x_2^{1/2} \over 2 \beta \delta_2^{3/2} x_1^{1/2}} + {15(\beta-1) x_2^{1/2} \over 4 \beta^2 \delta_2^{3/2} x_1^{3/2}}\right) + {e^{x_2} \lambda \over 1-x_2} \left( {9x_1^{3/2} x_2^{1/2}\over 2 \delta_2^{3/2} }+ {27 x_1^{1/2} x_2^{1/2} \over 4 \beta \delta_2^{3/2} } \right) + \cdots,\\
  \hat{C} \sim& e^{-x_1}  \left( {(\beta-1) x_1^{5/2} \over \beta} + {5 (\beta-1) x_1^{3/2} \over 2 \beta^2}\right) + \lambda \left(x_1^{9/2} + { 9 x_1^{7/2} \over 2 \beta} \right) + \cdots.
 \end{split}
 \label{extrema for two moduli}
\end{equation}
Plugging the extremal solution into the Hessian, we get, at the large volume limit
\begin{equation}
 \begin{split}
  \left(-{a_1^3 B_1 \, W_0\,  z \over 2}\right)^{-1}  \partial_{x_1}^2 V|_{\rm ext} \sim& e^{-x_1} \left({\beta-1 \over x_1^2} - {5(\beta-1) \over 2 x_1^3} \right) + \lambda \left(-{9 \beta \over 2 x_1} - {9 \over 2 x_1^2} \right) + \cdots,\\
  \left(-{a_1^3 B_1 \, W_0\,  z \over 2}\right)^{-1}  \partial_{y_1}^2 V_{\rm ext} \sim& e^{-x_1} \left(-{\beta-1 \over x_1^2} + {5 (\beta-1) \over 2 x_1^3} \right) + \lambda \left({9 \beta \over 2 x_1} + {45 \over 4 x_1^2} \right) + \cdots,\\
  \left(-{a_1^3 B_1 \, W_0\,  z \over 2}\right)^{-1}  \partial_{x_2}^2 V|_{\rm ext} \sim& e^{-x_1} {3 (\beta-1)(1+3 x_2 - 2 x_2^2) \over 4 \beta \delta_2^{3/2} x_1^{7/2} x_2^{1/2}(1-x_2)} + \lambda {9 (1+3 x_2 - 2 x_2^2) \over 4 \delta_2^{3/2} x_1^{3/2} x_2^{1/2} (1- x_2)} + \cdots,\\
  \left(-{a_1^3 B_1 \, W_0\,  z \over 2}\right)^{-1}  \partial_{x_1} \partial_{x_2} V|_{\rm ext} \sim& - e^{-x_1} {9(\beta-1) x_2^{1/2} \over 2 \beta \delta_2^{3/2} x_1^{9/2}} - \lambda {81 x_2^{1/2} \over 4 \beta \delta_2^{3/2} x_1^{7/2}} + \cdots,
 \end{split}
\end{equation}
where we keep the next-leading order terms in $\partial_{x_1}^2 V|_{\rm ext}, \partial_{y_1}^2 V|_{\rm ext}$ since their leading order terms may vanish due to the stability condition, while the leading order terms in 
 $\partial_{x_2}^2 V|_{\rm ext}, \partial_{x_1} \partial_{x_2} V|_{\rm ext}$ are not.

According to the Sylvester's criteria, the positivity of the sub-matrices are necessary conditions for the positivity of the entire matrix (see e.g. \cite{Gilbert:1991}, also applied for necessary stability constraints in \cite{Shiu:2011zt,VanRiet:2011yc,Danielsson:2012et}). 
So we consider the stability in the $x_1$-$y_1$ subspace first.
Similarly to the previous section, the condition in this subspace may be expressed as
\begin{equation}
 \begin{split}
  e^{-x_1} \left({2 (\beta-1) \over 9 \beta x_1} - {5 (\beta^2 -1) \over 9 \beta^2 x_1^2} + \cdots \right) \lesssim \lambda \lesssim e^{-x_1} \left({2 (\beta-1) \over 9 \beta x_1} - {5\beta^2- 3 \beta -2 \over 9 \beta^2 x_1^2 } + \cdots \right),
 \end{split}
 \label{leading lambda2}
\end{equation}
for large $x_1$ and $\beta>1$.
Recall that $d_2 >0$ is necessary for the stability on the $y_2$-direction, which becomes
\begin{equation}
 \begin{split}
  0< x_2 \lesssim 1, \quad - e^{-x_1}{ \beta-1 \over 3 \beta x_1^2} \lesssim \lambda, \qquad 
  {\rm or} \qquad 
  1\lesssim x_2, \quad \lambda \lesssim - e^{-x_1}{ \beta-1 \over 3 \beta x_1^2}.
 \end{split}
\end{equation}
So, to meet the condition (\ref{leading lambda2}), the solution is required to stay within $0<x_2 \lesssim 1$.

The remaining task is to check the stability condition in the $x_1$-$x_2$ subspace.
Now plugging the leading order of $\lambda$ into the Hessian, we get
\begin{equation}
 \begin{split}
  \left(-{a_1^3 B_1 \, W_0\,  z \over 2}\right)^{-1}  \partial_{x_1}^2 V_{\rm ext} \sim& e^{-x_1} {3 (\beta-1) \over 2 \beta x_1^3} + \cdots, \\
  \left(-{a_1^3 B_1 \, W_0\,  z \over 2}\right)^{-1}  \partial_{x_2}^2 V|_{\rm ext} \sim& e^{-x_1} {(\beta-1) (1+3x_2 - 2 x_2^2) \over 2 \beta \delta_2^{3/2} x_1^{5/2} x_2^{1/2} (1-x_2)} + \cdots,\\
  \left(-{a_1^3 B_1 \, W_0\,  z \over 2}\right)^{-1}  \partial_{x_1} \partial_{x_2} V_{\rm ext} \sim& - e^{-x_1} {9 (\beta^2-1) x_2^{1/2} \over 2 \beta^2 \delta_2^{3/2} x_1^{9/2}} + \cdots.
 \end{split}
\end{equation}
It is clear that the diagonal components are positive for $\beta>1, \ 0<x_2 <1$, while the off-diagonal components are sub-leading in the determinant; therefore, positivity of the Hessian is assured at the large volume limit.

Let us summarize the stability analysis in this section.
The extremal conditions for meta-stable vacua become
\begin{equation}
 \begin{split}
  {1\over z} \sim - {1\over \beta^3} e^{(\beta-1) x_1}, \quad
  d_2 \sim e^{-x_1+x_2} {(\beta-1) x_1^{1/2} x_2^{1/2} \over \beta \delta_2^{3/2} (1-x_2)}, \quad
  \hat{C} \sim e^{-x_1} {2 (\beta-1) x_1^{7/2} \over 9 \beta}, \quad
  \beta > 1.
 \end{split}
 \label{approx relation for two moduli}
\end{equation}
Then the solutions sit in the region
\begin{equation}
 1\ll x_1, \quad 0< x_2 < 1. 
\end{equation}
So having the racetrack type of potential for big volume modulus $x_1$ is well-motivated to realize the large volume even at $dS$ vacua.
Since we can easily have large volume solutions at $dS$ vacua, we can control the several stringy corrections, simultaneously realizing the cosmological constant which scales exponentially small:
\begin{equation}
 \begin{split}
  \left. \Lambda = V\right|_{\rm min} \sim {62\sqrt{2} a^{3/2} B^2 \beta^4 \over 243 \xi \sqrt{\beta-1}} \left(-\beta^{-3} z\right)^{2\beta \over \beta-1} \left(-\ln (-\beta^{-3} z)\right)^{5/2},
 \end{split}
\end{equation}
where we have used the approximate solution for $x_1$ using (\ref{approx relation for two moduli}), similar to (\ref{potmin at large volume}).

\begin{figure}
 \begin{center}
  \includegraphics[width=25em]{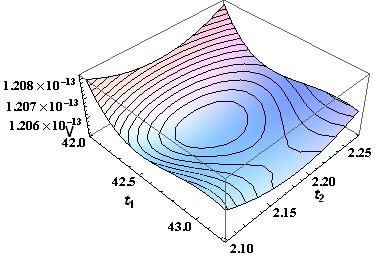}
 \end{center}
 \caption{
 A $dS$ minimum in the full potential (\ref{full-pot for two}).}
 \label{fig:multi-fullpot}
\end{figure}

We end this section by presenting a sample $dS$ solution.
In figure \ref{fig:multi-fullpot}, we illustrate a $dS$ solution in the full-potential (\ref{full-pot for two}) with the parameter set:
\begin{equation}
 \begin{split}
  &W_0 = -0.223, \ A_1 = 1.65, \ B_1 = - 4.77, \ A_2 = 2.24 \times 10^{-7},\\
  &a_1 = {2 \pi \over 15}, \ b_1 = {2 \pi \over 14}, \ a_2 = {2\pi \over 15}, \ \xi =4.35 \times 10^{-3}.
 \end{split}
\end{equation}
Then the solution is given by
\begin{equation}
 t_1 \sim 42.5, \quad t_2 \sim 2.18, \quad \tau_1 = \tau_2 =0. 
\end{equation}
So the large volume is easily realized: ${\cal V} \sim 774$.
At the local minimum of the potential,
\begin{equation}
 \Lambda=  1.21 \times 10^{-13}
\end{equation}
which is close to the previous result.

The Hessian is given as
\begin{equation}
 \begin{split}
  \partial_{t_1}^2 V|_{\rm min} =& 1.07 \times 10^{-15}, \quad
  \partial_{t_2}^2 V|_{\rm min} = 2.66 \times 10^{-14}, \quad
  \partial_{t_1} \partial_{t_2} V|_{\rm min} = -1.33 \times 10^{-16},\\
  \partial_{\tau_1}^2 V|_{\rm min} =& 3.79 \times 10^{-16}, \quad
  \partial_{\tau_2}^2 V|_{\rm min} = 2.14 \times 10^{-14}, \quad
  \partial_{\tau_1} \partial_{\tau_2} V|_{\rm min} = 1.17 \times 10^{-20}.
 \end{split}
\end{equation}
Therefore the minimum is stable in general.
There is no other cross term between the real and imaginary parts owing to $\tau_1=\tau_2=0$.

Note that the combined parameters here are given by
\begin{equation}
 \begin{split}
  &z = -0.346, \quad d_2 = 1.36 \times 10^{-7}, \quad \hat{C} = 1.06 \times 10^{-5}, \quad \beta \sim 1.07, \quad \delta_2 = 1.,\\
  &x_1 = 17.8, \quad x_2 = 0.915, \quad y_1 = y_2 = 0.
 \end{split}
\end{equation}

\section{Discussions \label{sec:discussions-}}

In this paper, the possibility of the large volume $x_1 \gg 1$ is allowed owing to the racetrack potential.
Since the volume is a parameter to control the $\alpha'$-corrections and also the string-loop corrections of order ${\cal O} ({\cal V}^{-10/3})$ in the potential \cite{Berg:2005ja,vonGersdorff:2005bf,Berg:2007wt,Cicoli:2007xp,Cicoli:2008va,Anguelova:2010ed} in light of the {\it extended no-scale structure} \cite{Berg:2007wt,Cicoli:2007xp}, the realization of the large volume is well-motivated to achieve a $dS$ vacuum which is meta-stable even in the presence of corrections.
The previous K\"ahler Uplift model is basically constrained by the dynamics, and suggests $x_1 \sim 3$.
Even though the large rank of the gauge group on D7-branes helps to relax the upper bound on the volume, we should be concerned with this way of relaxation due to the constraint of the gauge group rank by the D7-tadpole cancellation condition \cite{Cicoli:2011qg,Louis:2012nb}.
So relaxing the constraint for $x_1$ in the racetrack model is important for the construction of $dS$ vacua.

It is worth mentioning that the resultant cosmological constant is exponentially suppressed as a function of the volume.
At a $dS$ vacuum with a small cosmological constant, the corrections would be more suppressed owing to the large volume.
However, we should keep in mind that the combined parameter $\hat{C}$ is required to be exponentially small as well. $\hat{C}$ can be small due to an exponentially small $W_0$.
In fact, the peaked distribution of $W_0$ toward $W_0 = 0$ is obtained using a linear model for complex structure moduli \cite{Sumitomo:2012vx}.
We may expect that the sharper peaked distribution of $W_0$ is realized in the presence of more non-trivial couplings.
So the even exponentially small $W_0$ is quite conceivable.

Recently, a new $\alpha'$-correction is estimated for ${\cal N}=1$ compactification when the first Chern number of a three-dimensional K\"ahler base $B_3$ in M/F theory is non-trivial \cite{Grimm:2013gma}. Since the coefficient for this correction appears non-negligible, the volume is required to be large to suppress this correction \cite{Francisco:2013te}. Here, the important suppression parameter for this additional correction would be the ratio of the coefficients between ${\cal V}^{-10/3}$ term owing to the extended-no-scale structure \cite{Berg:2007wt,Cicoli:2007xp} and the leading $\alpha'$ correction term \cite{Becker:2002nn} scaling like ${\cal V}^{-3}$ in the potential.
Including this $\eta$ correction term to the potential $V$ (\ref{approximated potential}) we study in this paper, we would have (${\cal V} \sim x^{3/2}$),
$$\frac{\xi}{x^{9/2}} \rightarrow \frac{\xi}{x^{9/2}} -\frac{\eta}{x^5} = \frac{\xi_{g_s=1}}{x^{9/2} g_s^{3/2}}  -\frac{\eta}{x^5}, $$
where $\eta$ is the coefficient of  the new $\alpha'$-correction and the string coupling dependence of $\xi$ (\ref{fullpotential setup}) is made explicit.
Our qualitative result will remain valid if this remains positive for relatively large volume. This can be satisfied if either the volume is large or if
$\eta \lesssim \xi$. This later condition may be satisfied in the weak coupling approximation. Of course, the Racetrack K\"ahler Uplift model analyzed in this paper is applicable for large classes of models including those with trivial first Chern number of $B_3$, in which case $\eta=0$.

The probability distribution $P(\Lambda)$ is sharply peaked toward $\Lambda = 0$, explaining a natural statistical preference for a small cosmological constant.
The distribution $P(\Lambda)$ is diverging, but normalizable.
Since this mechanism works to have a hierarchical structure from Planck scale, we may worry about the cosmological moduli problem \cite{Coughlan:1983ci,Banks:1993en,deCarlos:1993jw}.
The cosmological moduli problem is a constraint from reheating of the universe, so it requires details of how the moduli fields decay to the matters.
Also there are some ways to relax the constraint, including a thermal inflation which is a mechanism to dilute the energy produced by the moduli coherent oscillation.
Recently, it is analyzed in detail that a double thermal inflation relaxes the constraint in the Large Volume Scenario \cite{Choi:2012ye}.
Therefore, the cosmological moduli issue crucially depends on the detail of cosmological history of the universe.

\section*{Acknowledgment}

We would like to thank Markus Rummel for stimulating discussions.

\appendix

\section{Case for non-zero axion value of the K\"ahler modulus \label{sec:case-non-zero}}

Here we like to show that the stability condition of the system with $y \neq 0$ at large volume cannot be satisfied. 
We first analyze the case for non-trivial imaginary mode $y\neq 0$ in the approximated potential (\ref{approximated potential}). Since $\partial_y V = 0$ is automatic for $y=0$, we shall assume $y \neq 0$.
The extremal condition $\partial_x V = \partial_y V = 0$ give the relations
\begin{equation}
 \begin{split}
  {1\over z} =& - e^{(\beta-1) x} { \sin y \over \beta^2 \sin (\beta y) }, \quad
  \hat{C} =  e^{-x} {x^{5/2} \over  \beta}\left({ \beta (x+2) \cos y} - {(\beta x + 2) \sin y \cot (\beta y) } \right).
 \end{split}
 \label{nonzero-y-sol}
\end{equation}
Substituting $z, \hat{C}$ in the potential, we get
\begin{equation}
 \begin{split}
  V|_{\rm ext} =& \left(-{a^3 B \, W_0\,  z \over 2}\right) e^{-x} {1\over 9 \beta x^2} \left( \beta { (2 x - 5)}   \cos y + {(-2 \beta x + 5)} {\sin y \cot (\beta y) }\right)\\
  \stackrel{x \gg 1}{\sim}&  \left(-{a^3 B \, W_0\,  z \over 2}\right) e^{-x_1} {2\over 9 x} \left( \cos y - {\sin y \cot (\beta y) } \right)+ \cdots
 \end{split}
\end{equation}
where in the last equation, we took the large volume approximation assuming $\beta \sim {\cal O} (1)$.

Next we consider the classical stability condition.
Again, using the relations (\ref{nonzero-y-sol}) and taking the large volume approximation for $V$ (\ref{approximated potential}), components of the Hessian at the extremal $\partial_i \partial_j V|_{\rm ext}$ are approximated by
\begin{equation}
 \begin{split}
  \partial_x^2 V|_{\rm ext} \sim& - \left(-{a^3 B \, W_0\,  z \over 2}\right) e^{-x} {1 \over x^2} \left( \cos y - {\sin y \cot (\beta y) } \right) + \cdots , \\
  \partial_y^2 V|_{\rm ext} =& \left(-{a^3 B \, W_0\,  z \over 2}\right) e^{-x} {1 \over x^2} \left( \cos y - {\sin y \cot (\beta y) } \right) ,\\
  \partial_x \partial_y V|_{\rm ext} =& \left(-{a^3 B \, W_0\,  z \over 2}\right) e^{-x}{\beta-1 \over x^2} \sin y.
 \end{split}
\end{equation}
Since the sign of the leading term of $\partial_x^2 V|_{\rm ext}$ is opposite to $\partial_y^2 V|_{\rm ext}$, we see that the leading term of the determinant is negative, i.e., $\det \left(\partial_i \partial_j V|_{\rm ext} \right) \lesssim 0$.
Note that the form of $\partial_y^2 V|_{\rm ext}$ is exact without the large volume approximation.
So no matter which cosmological constant we have, we cannot satisfy the stability condition for the system with $y \neq 0$ at the large volume.

\bibliographystyle{utphys}
\bibliography{C:/Users/lunatic/Dropbox/papers/myrefs}

\providecommand{\href}[2]{#2}\begingroup\raggedright\begin{thebibliography}{10}

\bibitem{Weinberg:1987dv}
S.~Weinberg, ``{Anthropic Bound on the Cosmological Constant},''
\href{http://dx.doi.org/10.1103/PhysRevLett.59.2607}{{\em Phys.Rev.Lett.} {\bf
  59} (1987)  2607}.

\bibitem{Bennett:2012fp}
C.~Bennett, D.~Larson, J.~Weiland, N.~Jarosik, G.~Hinshaw, {\em et al.},
  ``{Nine-Year Wilkinson Microwave Anisotropy Probe (WMAP) Observations: Final
  Maps and Results},''
\href{http://arxiv.org/abs/1212.5225}{{\tt arXiv:1212.5225 [astro-ph.CO]}}.

\bibitem{Riess:1998cb}
{\bf Supernova Search Team} Collaboration, A.~G. Riess {\em et al.},
  ``{Observational evidence from supernovae for an accelerating universe and a
  cosmological constant},'' \href{http://dx.doi.org/10.1086/300499}{{\em
  Astron.J.} {\bf 116} (1998)  1009--1038},
\href{http://arxiv.org/abs/astro-ph/9805201}{{\tt arXiv:astro-ph/9805201
  [astro-ph]}}.

\bibitem{Perlmutter:1998np}
{\bf Supernova Cosmology Project} Collaboration, S.~Perlmutter {\em et al.},
  ``{Measurements of Omega and Lambda from 42 high redshift supernovae},''
  \href{http://dx.doi.org/10.1086/307221}{{\em Astrophys.J.} {\bf 517} (1999)
  565--586},
\href{http://arxiv.org/abs/astro-ph/9812133}{{\tt arXiv:astro-ph/9812133
  [astro-ph]}}.

\bibitem{Douglas:2006es}
M.~R. Douglas and S.~Kachru, ``{Flux compactification},''
  \href{http://dx.doi.org/10.1103/RevModPhys.79.733}{{\em Rev.Mod.Phys.} {\bf
  79} (2007)  733--796},
\href{http://arxiv.org/abs/hep-th/0610102}{{\tt arXiv:hep-th/0610102}}.

\bibitem{Bousso:2000xa}
R.~Bousso and J.~Polchinski, ``{Quantization of four form fluxes and dynamical
  neutralization of the cosmological constant},'' {\em JHEP} {\bf 0006} (2000)
  006,
\href{http://arxiv.org/abs/hep-th/0004134}{{\tt arXiv:hep-th/0004134
  [hep-th]}}.

\bibitem{Sumitomo:2012wa}
Y.~Sumitomo and S.-H.~H. Tye, ``{A Stringy Mechanism for A Small Cosmological
  Constant},'' \href{http://dx.doi.org/10.1088/1475-7516/2012/08/032}{{\em
  JCAP} {\bf 1208} (2012)  032},
\href{http://arxiv.org/abs/1204.5177}{{\tt arXiv:1204.5177 [hep-th]}}.

\bibitem{Sumitomo:2012vx}
Y.~Sumitomo and S.~H. Tye, ``{A Stringy Mechanism for A Small Cosmological
  Constant - Multi-Moduli Cases -},''
  \href{http://dx.doi.org/10.1088/1475-7516/2013/02/006}{{\em JCAP} {\bf 1302}
  (2013)  006},
\href{http://arxiv.org/abs/1209.5086}{{\tt arXiv:1209.5086 [hep-th]}}.

\bibitem{Sumitomo:2012cf}
Y.~Sumitomo and S.-H.~H. Tye, ``{Preference for a Vanishingly Small
  Cosmological Constant in Supersymmetric Vacua in a Type IIB String Theory
  Model},''
\href{http://arxiv.org/abs/1211.6858}{{\tt arXiv:1211.6858 [hep-th]}}.

\bibitem{Balasubramanian:2004uy}
V.~Balasubramanian and P.~Berglund, ``{Stringy corrections to Kahler
  potentials, SUSY breaking, and the cosmological constant problem},''
  \href{http://dx.doi.org/10.1088/1126-6708/2004/11/085}{{\em JHEP} {\bf 0411}
  (2004)  085},
\href{http://arxiv.org/abs/hep-th/0408054}{{\tt arXiv:hep-th/0408054
  [hep-th]}}.

\bibitem{Westphal:2006tn}
A.~Westphal, ``{de Sitter string vacua from Kahler uplifting},''
  \href{http://dx.doi.org/10.1088/1126-6708/2007/03/102}{{\em JHEP} {\bf 0703}
  (2007)  102},
\href{http://arxiv.org/abs/hep-th/0611332}{{\tt arXiv:hep-th/0611332}}.

\bibitem{Rummel:2011cd}
M.~Rummel and A.~Westphal, ``{A sufficient condition for de Sitter vacua in
  type IIB string theory},''
  \href{http://dx.doi.org/10.1007/JHEP01(2012)020}{{\em JHEP} {\bf 1201} (2012)
   020},
\href{http://arxiv.org/abs/1107.2115}{{\tt arXiv:1107.2115 [hep-th]}}.

\bibitem{deAlwis:2011dp}
S.~de~Alwis and K.~Givens, ``{Physical Vacua in IIB Compactifications with a
  Single Kaehler Modulus},''
  \href{http://dx.doi.org/10.1007/JHEP10(2011)109}{{\em JHEP} {\bf 1110} (2011)
   109},
\href{http://arxiv.org/abs/1106.0759}{{\tt arXiv:1106.0759 [hep-th]}}.

\bibitem{Balasubramanian:2005zx}
V.~Balasubramanian, P.~Berglund, J.~P. Conlon, and F.~Quevedo, ``{Systematics
  of moduli stabilisation in Calabi-Yau flux compactifications},''
  \href{http://dx.doi.org/10.1088/1126-6708/2005/03/007}{{\em JHEP} {\bf 0503}
  (2005)  007},
\href{http://arxiv.org/abs/hep-th/0502058}{{\tt hep-th/0502058}}.

\bibitem{Berg:2005ja}
M.~Berg, M.~Haack, and B.~Kors, ``{String loop corrections to Kahler potentials
  in orientifolds},''
  \href{http://dx.doi.org/10.1088/1126-6708/2005/11/030}{{\em JHEP} {\bf 0511}
  (2005)  030},
\href{http://arxiv.org/abs/hep-th/0508043}{{\tt arXiv:hep-th/0508043
  [hep-th]}}.

\bibitem{vonGersdorff:2005bf}
G.~von Gersdorff and A.~Hebecker, ``{Kahler corrections for the volume modulus
  of flux compactifications},''
  \href{http://dx.doi.org/10.1016/j.physletb.2005.08.024}{{\em Phys.Lett.} {\bf
  B624} (2005)  270--274},
\href{http://arxiv.org/abs/hep-th/0507131}{{\tt arXiv:hep-th/0507131
  [hep-th]}}.

\bibitem{Berg:2007wt}
M.~Berg, M.~Haack, and E.~Pajer, ``{Jumping Through Loops: On Soft Terms from
  Large Volume Compactifications},''
  \href{http://dx.doi.org/10.1088/1126-6708/2007/09/031}{{\em JHEP} {\bf 0709}
  (2007)  031},
\href{http://arxiv.org/abs/0704.0737}{{\tt arXiv:0704.0737 [hep-th]}}.

\bibitem{Cicoli:2007xp}
M.~Cicoli, J.~P. Conlon, and F.~Quevedo,
  \href{http://dx.doi.org/10.1088/1126-6708/2008/01/052}{``{Systematics of
  String Loop Corrections in Type IIB Calabi-Yau Flux Compactifications},''{\em
  JHEP} {\bf 0801} (Aug., 2008)  052},
\href{http://arxiv.org/abs/0708.1873}{{\tt 0708.1873}}.

\bibitem{Cicoli:2008va}
M.~Cicoli, J.~P. Conlon, and F.~Quevedo, ``{General Analysis of LARGE Volume
  Scenarios with String Loop Moduli Stabilisation},''
  \href{http://dx.doi.org/10.1088/1126-6708/2008/10/105}{{\em JHEP} {\bf 0810}
  (2008)  105},
\href{http://arxiv.org/abs/0805.1029}{{\tt arXiv:0805.1029 [hep-th]}}.

\bibitem{Anguelova:2010ed}
L.~Anguelova, C.~Quigley, and S.~Sethi, ``{The Leading Quantum Corrections to
  Stringy Kahler Potentials},''
  \href{http://dx.doi.org/10.1007/JHEP10(2010)065}{{\em JHEP} {\bf 1010} (2010)
   065},
\href{http://arxiv.org/abs/1007.4793}{{\tt arXiv:1007.4793 [hep-th]}}.

\bibitem{Westphal:2005yz}
A.~Westphal, ``{Eternal inflation with alpha-prime-corrections},''
  \href{http://dx.doi.org/10.1088/1475-7516/2005/11/003}{{\em JCAP} {\bf 0511}
  (2005)  003},
\href{http://arxiv.org/abs/hep-th/0507079}{{\tt arXiv:hep-th/0507079
  [hep-th]}}.

\bibitem{Chen:2011ac}
X.~Chen, G.~Shiu, Y.~Sumitomo, and S.~H. Tye, ``{A Global View on The Search
  for de-Sitter Vacua in (type IIA) String Theory},''
  \href{http://dx.doi.org/10.1007/JHEP04(2012)026}{{\em JHEP} {\bf 1204} (2012)
   026},
\href{http://arxiv.org/abs/1112.3338}{{\tt arXiv:1112.3338 [hep-th]}}.

\bibitem{Bachlechner:2012at}
T.~C. Bachlechner, D.~Marsh, L.~McAllister, and T.~Wrase, ``{Supersymmetric
  Vacua in Random Supergravity},''
  \href{http://dx.doi.org/10.1007/JHEP01(2013)136}{{\em JHEP} {\bf 1301} (2013)
   136},
\href{http://arxiv.org/abs/1207.2763}{{\tt arXiv:1207.2763 [hep-th]}}.

\bibitem{Marsh:2011aa}
D.~Marsh, L.~McAllister, and T.~Wrase, ``{The Wasteland of Random
  Supergravities},'' \href{http://dx.doi.org/10.1007/JHEP03(2012)102}{{\em
  JHEP} {\bf 1203} (2012)  102},
\href{http://arxiv.org/abs/1112.3034}{{\tt 1112.3034}}.

\bibitem{Becker:2002nn}
K.~Becker, M.~Becker, M.~Haack, and J.~Louis, ``{Supersymmetry breaking and
  alpha-prime corrections to flux induced potentials},'' {\em JHEP} {\bf 0206}
  (2002)  060,
\href{http://arxiv.org/abs/hep-th/0204254}{{\tt hep-th/0204254}}.

\bibitem{Kachru:2003aw}
S.~Kachru, R.~Kallosh, A.~D. Linde, and S.~P. Trivedi, ``{De Sitter vacua in
  string theory},'' \href{http://dx.doi.org/10.1103/PhysRevD.68.046005}{{\em
  Phys.Rev.} {\bf D68} (2003)  046005},
\href{http://arxiv.org/abs/hep-th/0301240}{{\tt arXiv:hep-th/0301240}}.

\bibitem{Escoda:2003fa}
C.~Escoda, M.~Gomez-Reino, and F.~Quevedo, ``{Saltatory de Sitter string
  vacua},'' {\em JHEP} {\bf 0311} (2003)  065,
\href{http://arxiv.org/abs/hep-th/0307160}{{\tt arXiv:hep-th/0307160
  [hep-th]}}.

\bibitem{BlancoPillado:2004ns}
J.~Blanco-Pillado, C.~Burgess, J.~M. Cline, C.~Escoda, M.~Gomez-Reino, {\em et
  al.}, ``{Racetrack inflation},''
  \href{http://dx.doi.org/10.1088/1126-6708/2004/11/063}{{\em JHEP} {\bf 0411}
  (2004)  063},
\href{http://arxiv.org/abs/hep-th/0406230}{{\tt arXiv:hep-th/0406230
  [hep-th]}}.

\bibitem{Kallosh:2004yh}
R.~Kallosh and A.~D. Linde, ``{Landscape, the scale of SUSY breaking, and
  inflation},'' \href{http://dx.doi.org/10.1088/1126-6708/2004/12/004}{{\em
  JHEP} {\bf 0412} (2004)  004},
\href{http://arxiv.org/abs/hep-th/0411011}{{\tt arXiv:hep-th/0411011}}.

\bibitem{Kachru:2002gs}
S.~Kachru, J.~Pearson, and H.~L. Verlinde, ``{Brane / flux annihilation and the
  string dual of a nonsupersymmetric field theory},'' {\em JHEP} {\bf 0206}
  (2002)  021,
\href{http://arxiv.org/abs/hep-th/0112197}{{\tt arXiv:hep-th/0112197}}.

\bibitem{Louis:2012nb}
J.~Louis, M.~Rummel, R.~Valandro, and A.~Westphal, ``{Building an explicit de
  Sitter},'' \href{http://dx.doi.org/10.1007/JHEP10(2012)163}{{\em JHEP} {\bf
  1210} (2012)  163},
\href{http://arxiv.org/abs/1208.3208}{{\tt arXiv:1208.3208 [hep-th]}}.

\bibitem{Collinucci:2008pf}
A.~Collinucci, F.~Denef, and M.~Esole, ``{D-brane Deconstructions in IIB
  Orientifolds},'' \href{http://dx.doi.org/10.1088/1126-6708/2009/02/005}{{\em
  JHEP} {\bf 0902} (2009)  005},
\href{http://arxiv.org/abs/0805.1573}{{\tt arXiv:0805.1573 [hep-th]}}.

\bibitem{Cicoli:2011qg}
M.~Cicoli, C.~Mayrhofer, and R.~Valandro, ``{Moduli Stabilisation for Chiral
  Global Models},'' \href{http://dx.doi.org/10.1007/JHEP02(2012)062}{{\em JHEP}
  {\bf 1202} (2012)  062},
\href{http://arxiv.org/abs/1110.3333}{{\tt arXiv:1110.3333 [hep-th]}}.

\bibitem{MartinezPedrera:2012rs}
D.~Martinez-Pedrera, D.~Mehta, M.~Rummel, and A.~Westphal, ``{Finding all flux
  vacua in an explicit example},''
\href{http://arxiv.org/abs/1212.4530}{{\tt arXiv:1212.4530 [hep-th]}}.

\bibitem{Danielsson:2012by}
U.~Danielsson and G.~Dibitetto, ``{On the distribution of stable de Sitter
  vacua},'' \href{http://dx.doi.org/10.1007/JHEP03(2013)018}{{\em JHEP} {\bf
  1303} (2013)  018},
\href{http://arxiv.org/abs/1212.4984}{{\tt arXiv:1212.4984 [hep-th]}}.

\bibitem{Ashok:2003gk}
S.~Ashok and M.~R. Douglas, ``{Counting flux vacua},''
  \href{http://dx.doi.org/10.1088/1126-6708/2004/01/060}{{\em JHEP} {\bf 0401}
  (2004)  060},
\href{http://arxiv.org/abs/hep-th/0307049}{{\tt arXiv:hep-th/0307049
  [hep-th]}}.

\bibitem{Denef:2004ze}
F.~Denef and M.~R. Douglas, ``{Distributions of flux vacua},''
  \href{http://dx.doi.org/10.1088/1126-6708/2004/05/072}{{\em JHEP} {\bf 0405}
  (2004)  072},
\href{http://arxiv.org/abs/hep-th/0404116}{{\tt arXiv:hep-th/0404116}}.

\bibitem{Denef:2004cf}
F.~Denef and M.~R. Douglas, ``{Distributions of nonsupersymmetric flux
  vacua},'' \href{http://dx.doi.org/10.1088/1126-6708/2005/03/061}{{\em JHEP}
  {\bf 0503} (2005)  061},
\href{http://arxiv.org/abs/hep-th/0411183}{{\tt arXiv:hep-th/0411183}}.

\bibitem{Denef:2004dm}
F.~Denef, M.~R. Douglas, and B.~Florea, ``{Building a better racetrack},''
  \href{http://dx.doi.org/10.1088/1126-6708/2004/06/034}{{\em JHEP} {\bf 0406}
  (2004)  034},
\href{http://arxiv.org/abs/hep-th/0404257}{{\tt hep-th/0404257}}.

\bibitem{Gray:2012jy}
J.~Gray, Y.-H. He, V.~Jejjala, B.~Jurke, B.~D. Nelson, {\em et al.},
  ``{Calabi-Yau Manifolds with Large Volume Vacua},''
  \href{http://dx.doi.org/10.1103/PhysRevD.86.101901}{{\em Phys.Rev.} {\bf D86}
  (2012)  101901},
\href{http://arxiv.org/abs/1207.5801}{{\tt arXiv:1207.5801 [hep-th]}}.

\bibitem{Gilbert:1991}
G.~T. Gilbert, ``Positive definite matrices and sylvester's criterion,'' {\em
  The American Mathematical Monthly} {\bf 98} (1991) no.~1, pp. 44--46.
  \url{http://www.jstor.org/stable/2324036}.

\bibitem{Shiu:2011zt}
G.~Shiu and Y.~Sumitomo, ``{Stability Constraints on Classical de Sitter
  Vacua},'' \href{http://dx.doi.org/10.1007/JHEP09(2011)052}{{\em JHEP} {\bf
  1109} (2011)  052},
\href{http://arxiv.org/abs/1107.2925}{{\tt arXiv:1107.2925 [hep-th]}}.

\bibitem{VanRiet:2011yc}
T.~Van~Riet, ``{On classical de Sitter solutions in higher dimensions},''
  \href{http://dx.doi.org/10.1088/0264-9381/29/5/055001}{{\em
  Class.Quant.Grav.} {\bf 29} (2012)  055001},
\href{http://arxiv.org/abs/1111.3154}{{\tt arXiv:1111.3154 [hep-th]}}.

\bibitem{Danielsson:2012et}
U.~H. Danielsson, G.~Shiu, T.~Van~Riet, and T.~Wrase, ``{A note on obstinate
  tachyons in classical dS solutions},''
  \href{http://dx.doi.org/10.1007/JHEP03(2013)138}{{\em JHEP} {\bf 1303} (2013)
   138},
\href{http://arxiv.org/abs/1212.5178}{{\tt arXiv:1212.5178 [hep-th]}}.

\bibitem{Grimm:2013gma}
T.~W. Grimm, R.~Savelli, and M.~Weissenbacher, ``{On $\alpha'$ corrections in
  N=1 F-theory compactifications},''
\href{http://arxiv.org/abs/1303.3317}{{\tt arXiv:1303.3317 [hep-th]}}.

\bibitem{Francisco:2013te}
F.~Pedro, M.~Rummel, and W.~Alexander, ``$\alpha'^2$ corrections and the
  extended no-scale structure,'' {\em to appear}  .

\bibitem{Coughlan:1983ci}
G.~Coughlan, W.~Fischler, E.~W. Kolb, S.~Raby, and G.~G. Ross, ``{Cosmological
  Problems for the Polonyi Potential},''
\href{http://dx.doi.org/10.1016/0370-2693(83)91091-2}{{\em Phys.Lett.} {\bf
  B131} (1983)  59}.

\bibitem{Banks:1993en}
T.~Banks, D.~B. Kaplan, and A.~E. Nelson, ``{Cosmological implications of
  dynamical supersymmetry breaking},''
  \href{http://dx.doi.org/10.1103/PhysRevD.49.779}{{\em Phys.Rev.} {\bf D49}
  (1994)  779--787},
\href{http://arxiv.org/abs/hep-ph/9308292}{{\tt hep-ph/9308292}}.

\bibitem{deCarlos:1993jw}
B.~de~Carlos, J.~Casas, F.~Quevedo, and E.~Roulet, ``{Model independent
  properties and cosmological implications of the dilaton and moduli sectors of
  4-d strings},'' \href{http://dx.doi.org/10.1016/0370-2693(93)91538-X}{{\em
  Phys.Lett.} {\bf B318} (1993)  447--456},
\href{http://arxiv.org/abs/hep-ph/9308325}{{\tt hep-ph/9308325}}.

\bibitem{Choi:2012ye}
K.~Choi, W.-I. Park, and C.~S. Shin, ``{Cosmological moduli problem in large
  volume scenario and thermal inflation},''
  \href{http://dx.doi.org/10.1088/1475-7516/2013/03/011}{{\em JCAP} {\bf 1303}
  (2013)  011},
\href{http://arxiv.org/abs/1211.3755}{{\tt arXiv:1211.3755 [hep-ph]}}.

\end{thebibliography}\endgroup

\end{document}